# Were the "Critical Evidence" presented in the South Korean Official Cheonan Report fabricated?

Seung-Hun Lee[1] and Panseok Yang[2]
[1]Department of Physics, University of Virginia, Charlottesville, VA 22904, UVA
[2]Department of Geological Sciences, University of Manitoba, Winnipeg, Manitoba, R3T 2N2, Canada

**South Korean Joint Investigation Group (JIG) [1] presented two "critical scientific evidence" that link the sinking of the South Korean navy corvette Cheonan on March 26, 2010 to the alleged explosion of a North Korean torpedo: the now-infamous "No. 1" blue ink mark on the torpedo, and the electron-dispersive spectroscopy (EDS) and x-ray diffraction (XRD) data of the three "adsorbed materials" extracted from the ship, the torpedo and a small-scale test-explosion. In our previous paper [2], we described the inconsistency of JIG's EDS and XRD data. Here we report our SEM, EDS, and XRD analyses on an Al powder that underwent melting followed by rapid quenching, and our simulation of the EDS spectra of $Al_2O_3$ and $Al(OH)_3$. We obtained an experimental value of 0.25 for the EDS intensity ratio, I(O)/I(Al), and a simulation result of I(O)/I(Al) ~ 0.23 for $Al_2O_3$ formed on the surface of the adsorbed materials during an explosion. The JIG's EDS data, however, showed very different values of ~ 0.9 for the same ratio. Interestingly, the high value of ~ 0.9 is expected for aluminum hydroxide, such as $Al(OH)_3$. Our results indicate that the JIG's adsorbed materials taken from the ship and torpedo are not associated with any explosion, and that the JIG's EDS data of their test-explosion sample are most likely fabricated. Until the South Korean JIG can convince the international scientific communities the validity of their data, their official conclusion that the torpedo sank the Cheonan ship should be discarded.**

Let us start with the key claims made in the Cheonan report and our main conclusions.

**Key claims made in the CheonAn report by the South Korean JIG:**

(a) Chemical and structural properties of the following three samples of "adsorbed material" (AM) were investigated using EDS and XRD: (i) the first sample, here referred to as AM-I, was extracted from the surface of the bow, stern, stack of the CheonAn ship; (ii) the second sample, AM-II, was extracted from the surface of the propeller of the torpedo; (iii) the third sample, AM-III, was from the inner surface of the Al plate used to cover the top of the metal tank that housed ocean water and used a small scale test explosion.

(b) The sample AM-III that was attached to the Al plate after the test explosion must be explosive-turned material.

(c) The EDS data of all three samples were almost identical, with similar strengths of aluminum and oxygen signals: with I(O)/I(Al) ~ 0.9. This confirms that the first two samples, AM-I and AM-II, are also explosive-turned material.

(d) However, the XRD patterns of the AM-I and AM-II samples taken from the CheonAn ship and from the torpedo did not exhibit any signature of aluminum or any oxidized Al material.

(e) The absence of any significant Al-related x-ray signals in the two samples, AM-I and AM-II, is in fact pivotal evidence that the torpedo exploded and sank the CheonAn ship. It is because during the explosion 100 % of the aluminum melted and supercooled into an amorphous Al-oxide, and an amorphous material does not produce a distinct X-ray diffraction pattern.

(f) The "adsorbed materials" that was attached on the propeller of the torpedo and on the CheonAn ship were determined to be identical. This leads to our conclusion that the torpedo explosion sank the corvette CheonAn.

**Summary of our SEM, EDS and XRD data on a heated and quenched Al sample, and simulations of EDS spectra of $Al_2O_3$ and $Al(OH)_3$**

(a) Their EDS and x-ray data are self-contradicting. Furthermore, the EDS I(O)/I(Al) ratio of their three samples, ~ 0.9, tells us that the adsorbed materials taken from the ship and torpedo are not associated with any explosion.

(b) We have performed SEM, EDS, and XRD measurements on two crystalline Al samples, one without heat treatment and another with heat and quenching treatment. Our results show that during heating and rapid quenching Al oxidizes only partially, and furthermore the resulting Al and $Al_2O_3$ are crystalline rather than amorphous. Even if 100 % Al turned into 100% amorphous $Al_2O_3$, it should have been observed in their XRD pattern.

(c) We also performed our simulation of the EDS spectra of $Al_2O_3$ and $Al(OH)_3$. Our simulation results yield I(O)/I(Al) ~ 0.23 for $Al_2O_3$ and ~ 0.85 for $Al(OH)_3$. Thus, we conclude that the JIG's AM-I and AM-II contain mainly $Al(OH)_3$.

**(d)** The high I(O)/I(Al) ratio of 0.81 for the AM-III reported by JIG was most likely fabricated in order to claim that the AM-I and AM-II samples taken from the ship and the torpedo are explosive-related materials.

**(e)** The JIG is not releasing their AM-III sample. Since the sample was obtained from a test experiment, we suggest re-performing the test explosion experiment, using a cover plate different than aluminum as was in the JIG experiment

**Our SEM, EDS and XRD Data and Discussion**

In order to distinguish the two aforementioned scenarios, we have performed SEM, EDS and XRD measurements on the following two samples: (1) Al powder without heat treatment and (2) an Al powder that was heated at 1100 °C for 40 min and quenched in water in less than 2 sec.

Here we briefly describe the experimental procedures.

A very fine powder of crystalline Al with 300 mesh, i.e., a maximum grain size of ~ 1 inch/300 ~ 85 $\mu$m, was chosen to maximize the oxidization during the experiment. We prepared two Al samples. Before the measurements, one sample, let say Al-a, was just taken out of the sample bottle without any heat treatment, while the other sample, Al-b, was put in a horizontal furnace, and heat up to 1100 °C that is well above the melting temperature of Al, 660 °C, and stayed at that 1100 °C for ~ 40 minutes, before it was taken out quickly and put into a jar of water. The procedure of taking out and putting it into the water took less than 2 seconds.

Each sample was divided into two portions: a larger amount that was used for the XRD and a smaller amount used for the SEM and EDS.

The x-ray portion was placed on a glass plate, while the EDS portion was placed on a carbon tape that was attached to a round carbon plate. The pictures of the x-ray and the EDS machine, and the loading of the samples were shown in the Supplementary Information. One thing to remember is that x-ray penetrates into the sample about a few hundreds $\mu$m while EDS penetrates only about several $\mu$m. Thus, EDS probes the atomic composition of the surface of the sample while x-ray probes the chemical material composition of the sample in bulk.

Fig. 1 (a)-(c) show the SEM images and the EDS data obtained from the first sample, Al-a. Fig. 1 (a) shows the SEM image, exhibiting the Al-a grains with typical sizes of ~ 10 $\mu$m to ~ 100 $\mu$m. Fig. 1 (b) is a close-up of one large grain whose EDS data is shown in Fig. 1 (c). The EDS data of Al-a clearly exhibits the Al peak at around 1.5 keV. And there are no other peaks, except the signal of the carbon coated to give electrical conductivity for the measurements. Other grains also gave similar EDS data. On the other hand, its x-ray data (Fig. 2 (a)) exhibit only the pure Al Bragg peaks, which indicates that as expected the amount of $Al_2O_3$ is negligible in this untreated crystalline Al powder sample.

Fig. 1 (d)-(f) show the SEM images and the EDS data of the melted and then quenched second sample, Al-b. The SEM images (Fig. 1 (d) and (e)) are much shinier than those of the Al-a sample (Fig. 1 (a) and (b)). This tells that the surfaces of the metallic Al grains are oxidized into the insulating $Al_2O_3$ from which the incident electrons are reflected rather than penetrating into the sample. Thus, the shiny SEM images indicate that oxidization occurred during the heating and quenching.

Now, the questions that arise are (1) how much Al has been oxidized during the heat treatment, and (2) in which form, crystalline or amorphous, the resulting oxidized Al and/or the remaining Al exist afterward. EDS and XRD data will give us the answers.

As shown in Fig. 2 (b), the x-ray data of the heat-treated sample, Al-b, exhibit two sets of well-defined sharp Bragg peaks from crystalline Al and crystalline $Al_2O_3$. This clearly indicates that **(1)** Al ***oxidizes partially*** not entirely during the heating and rapid quenching, and that (2) after the quenching the Al and the oxidized $Al_2O_3$ become ***crystalline*** rather than amorphous! Our refinement of the data tells us that ~ 40 % of the Al has oxidized during the heat treatment. When compared with the x-ray data of the test-explosion sample, AM-III, presented in the official Cheonan report, in the test-explosion the amount of oxidized Al is much smaller than the 40 % oxidization observed here. We believe that it is because in our experiment the Al sample was above its melting temperature for ~ 40 minutes that is much longer than it would have during an explosion. Of course, the amount of the oxidized Al would depends on other factors as well such as the amount of oxygen in the surroundings and temperature, but from our heat-treatment results and their test-explosion results, one can safely say that the duration of the sample above its melting point is the most crucial, and the 40 % is probably close to the maximum oxidization fraction for an explosion. This was proved by their test-explosion.[1]

What are the implications of our results to the EDS and XRD data of the three "adsorbed materials" that were presented as one of the two "critical scientific evidence" in the South Korean official Cheonan report? **(1)** Our XRD data of the heat-treated sample, Al-b, are similar to those of the "adsorbed material", AM-III, taken from their test-explosion [1,2]: strong Bragg peaks from crystalline Al and much weaker Bragg peaks from crystalline $Al_2O_3$ were present. This suggests that the JIG's XRD data of the AM-III is correct, on the contrary to the JIG's claim that the Al Bragg peaks observed in their AM-III XRD data are not associated with explosives because the XRD measurements were done with the Al-cover plate to which the AM-III samples were attached. **(2)** In the EDS data, our AM-b sample and the JIG's AM-III sample both exhibit Al and O signals. However, the relative strength of the Al and O EDS signals, I(O)/I(Al), is very different in the two samples: I(O)/I(Al) = 0.25 for our AM-b (Fig. 1 (f)) while I(O)/I(Al) = 0.81 for JIG's AM-III sample (Fig. 3 (c)).

In order to understand the discrepancy in the I(O)/I(Al) ratio, we simulated the

EDS spectra for $Al_2O_3$ using the EDS simulation software, NIST DTSA II. As shown in Fig. 4, the I(O)/I(Al) ratio should be ~ 0.23 for surface of samples that are predominantly $Al_2O_3$ (red line), as is the case in an explosion. This simulation result is consistent with our EDS data, I(O)/I(Al) ~ 0.25 (Fig. 1 (f)). Other previous experimental studies have also shown similar value for the I(O)/I(Al) ratio.[3] On the other hand, as seen in Fig. 3, all three samples, AM-I, AM-II, AM-III, presented by the JIG as evidence showed very high I(O)/I(Al) ratios, 0.92 (AM-I), 0.90 (AM-II), and 0.81 (AM-III).[1] This is strikingly different from our experimental and simulated result and previously measured I(O)/I(Al) ratio. What kind of Al-related materials would yield such a high I(O)/I(Al) ratio? Fig. 4 shows the simulated EDS spectra for gibbsite ($Al(OH)_3$) (black line) that is a clay mineral commonly formed by weathering processes, and it clearly shows that I(O)/I(Al) ~ 0.85 for the gibbsite. Thus, if the JIG's EDS data for their AM-I and AM-II samples are correct, then their adsorbed materials are not associated with an explosive. Instead, the EDS spectra of AM-I and AM-II are mainly hydrated alumina, such as gibbsite $Al(OH)_3$.

One may ask if the aluminum hydroxides can form during an explosion. However, several scientific experiments that studied the environment of a real explosion, have reported that the resulting Al-related compounds are crystalline aluminum, crystalline $\alpha$-$Al_2O_3$, and amorphous $\gamma$-$Al_2O_3$.[4,5,6] Thus, the JIG's adsorbed materials AM-I and AM-II have nothing to do with any explosion, and cannot be used as an evidence for the alleged explosion of the torpedo. During their test explosion experiment, there was insufficient time for Al to react with water to form the gibbsite, $Al(OH)_3$, indicating that the high I(O)/I(Al) ratio of 0.81 for the AM-III reported by JIG was fabricated in order to claim that the AM-I and AM-II samples taken from the ship and the torpedo are explosive-related materials.

When South Korean congresswoman, Lee, JungHee, recently asked the JIG to release their adsorbed samples, they released only the AM-I and AM-II. According to JIG, the reason for not releasing their AM-III sample was that they used up all the AM-III samples. First, EDS and XRD measurements use tiny samples, a few tenth of a milligram and a few milligrams, respectively (see Fig. 5). Second, EDS and XRD techniques are non-destructive. Thus, the JIG's inability to release any AM-III sample indicates either a complete mishandling of the samples during testing, or simply intentional hiding of their AM-III sample. As shown in Fig. 6, the JIG's test explosion experiment produced several spots of the sample on the Al-cover plate, which contain more than sufficient amount of materials for several EDS and XRD measurements. Their reluctance to provide AM-III indicates that the JIG is intentionally hiding or has destroyed their AM-III sample. Since the sample was obtained from a test experiment, one can simply redo the experiment. We suggest re-performing the test explosion experiment with the following conditions:

1. Before the experiment, take ~ 0.1g from the explosive, and run EDS and XRD analyses to make sure that the explosive contains polycrystalline aluminum.
2. After the explosion, extract adsorbed materials from the cover plate, and perform EDS and x-ray measurements in proper manner.

In addition, we suggest using a cover plate different than aluminum as was in the JIG experiment to prevent contamination.

**Acknowlegement**

We thank Representative MoonSoon Choi for providing the information, and a few people for discussion and for help with the SEM, EDS and x-ray measurements.

**Figures**

Fig. 1. SEM and EDS data of (a)-(c) the untreated Al sample and (d)-(f) the melted and rapidly quenched Al sample. (a), (b), (d) and (e) are the SEM images. (c) and (f) are the EDS data.

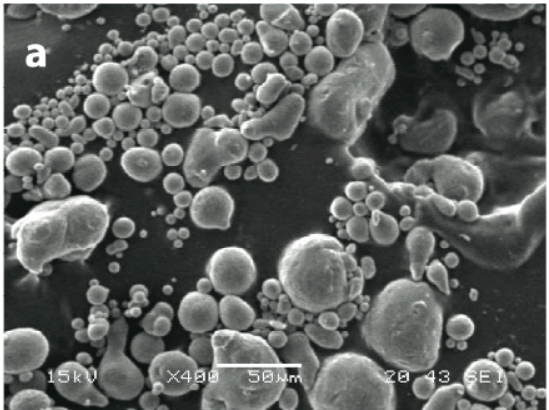
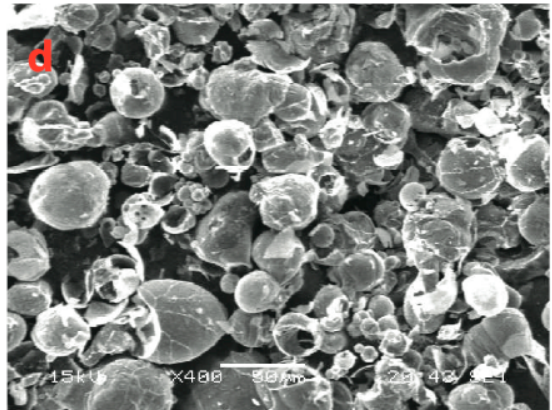
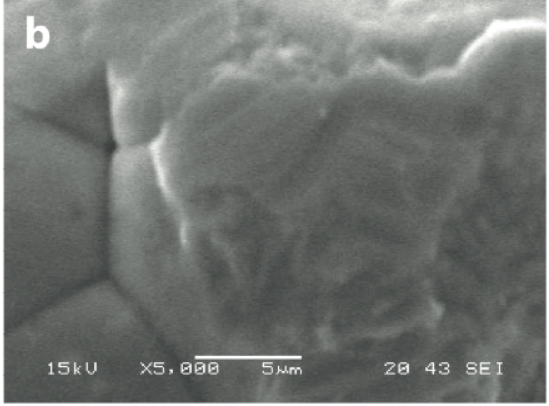
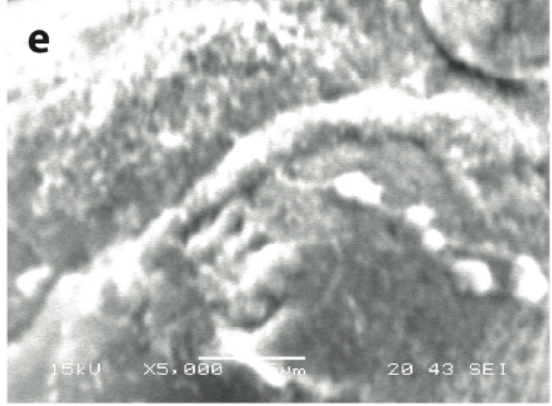
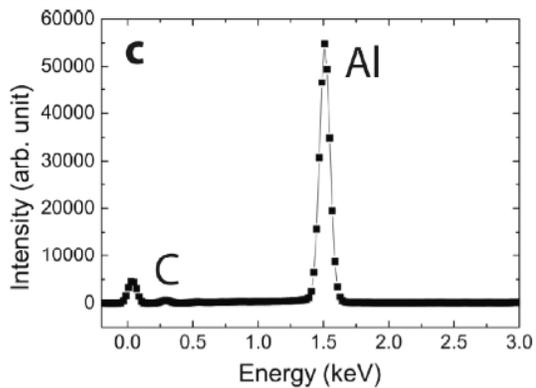
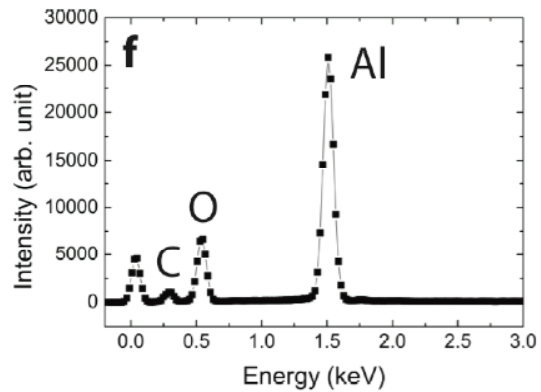

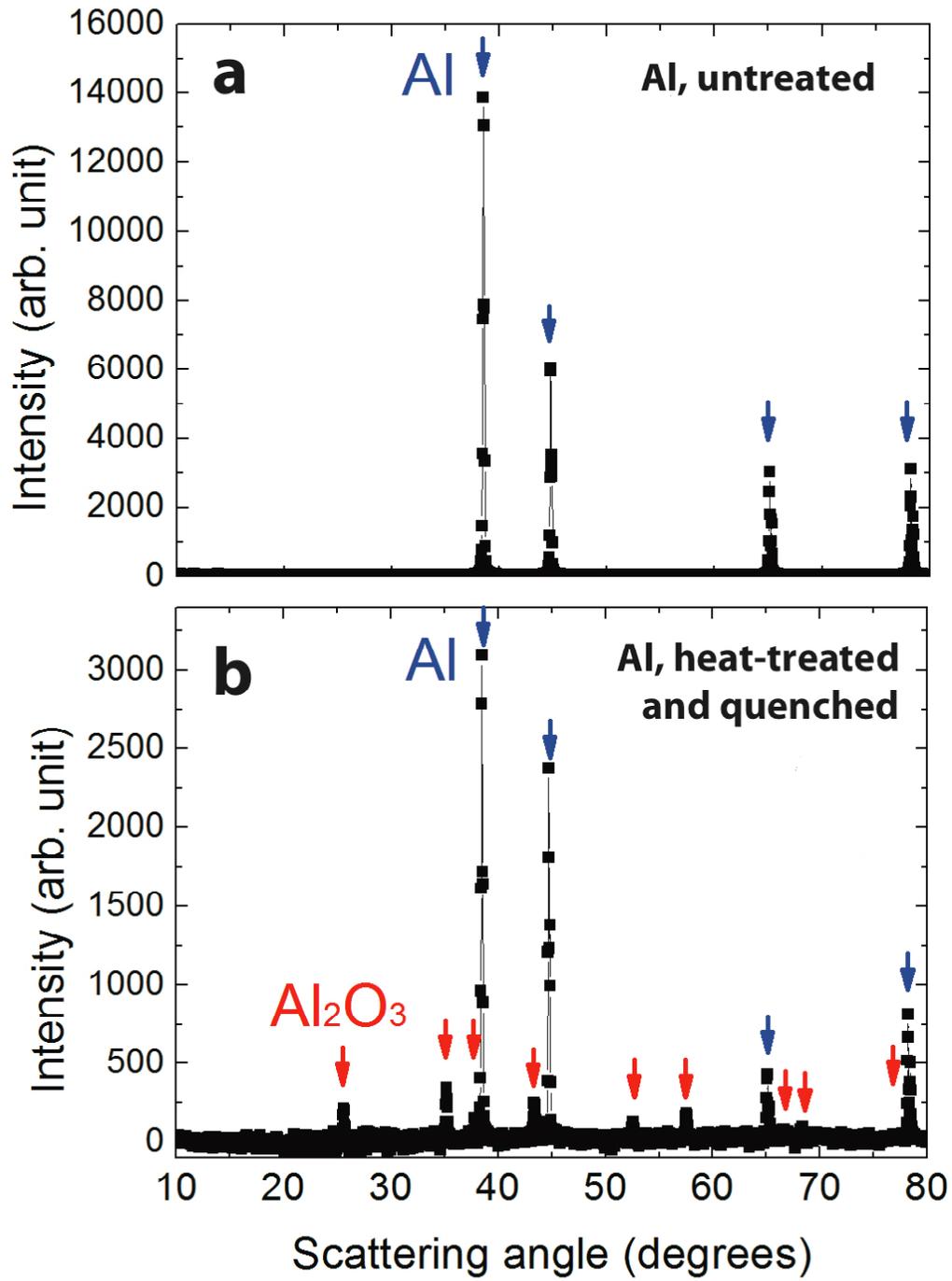

Fig. 2. X-ray data of (a) the untreated Al sample and (b) the melted and rapidly quenched Al sample.

Fig. 3. The reported EDS data obtained from (a) AM-I (left panel), the absorbed material extracted from the "bow, stern, and stack" of the Cheon-An ship, (b) from AM-II (center), "critical evidence", extracted from the propeller of the torpedo, and (c) from AM-III (right panel), "UNDEX Experiment", extracted from the mock-up torpedo after explosion. The horizontal and vertical axes represent the energy and strength of the signal, respectively. This figure was taken from Ref. [1].

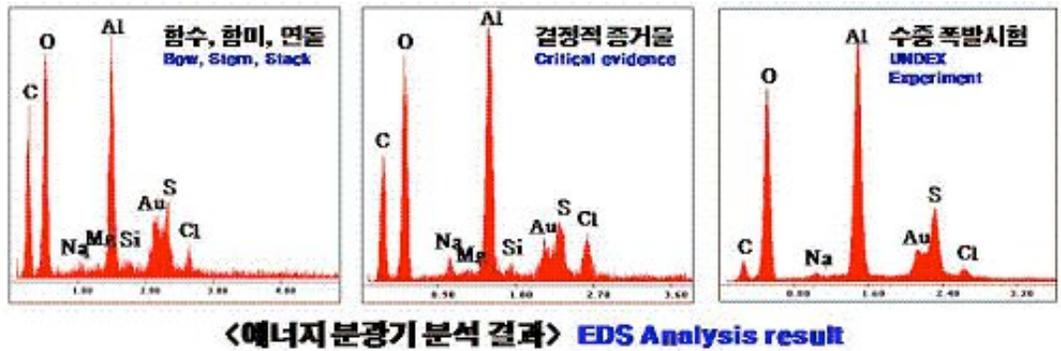

Fig. 4. EDS simulation results for $Al_2O_3$ (red line) and $Al(OH)_3$ (black line).

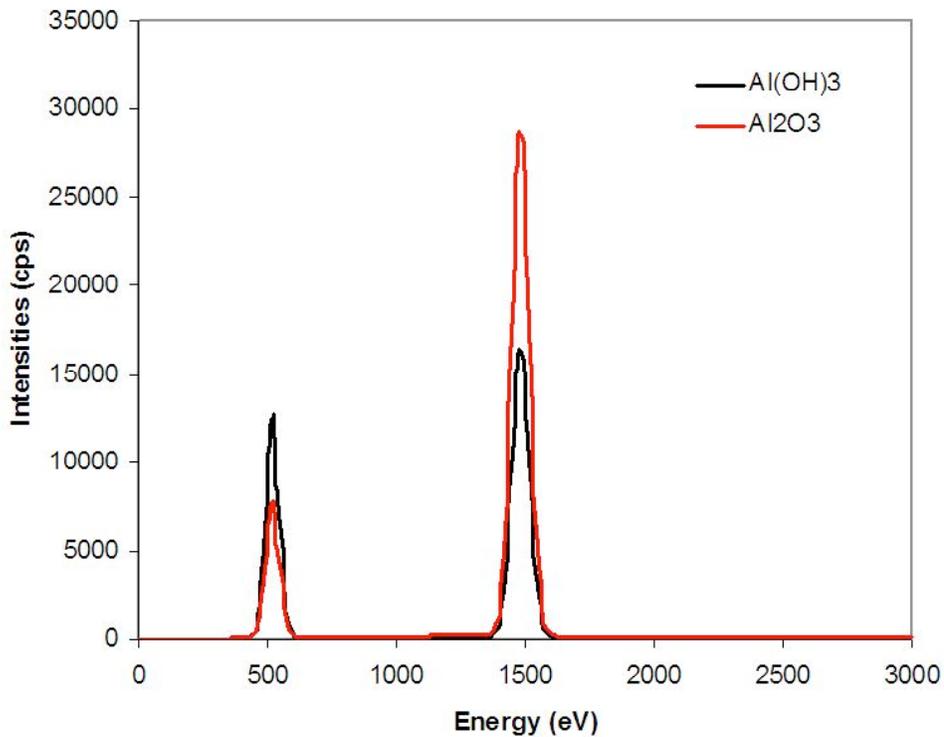

Fig. 5. Typical samples for (a) XRD and (b) EDS measurements. The plastic ruler shows length in red marks in the unit of cm. In (a), a thin layer of the white sample of 4.3 milligrams is placed on a glass plate. In (b), a white spot of 0.3 milligram sample is placed on a round carbon plate.

(a) x-ray sample

(b) EDS sample

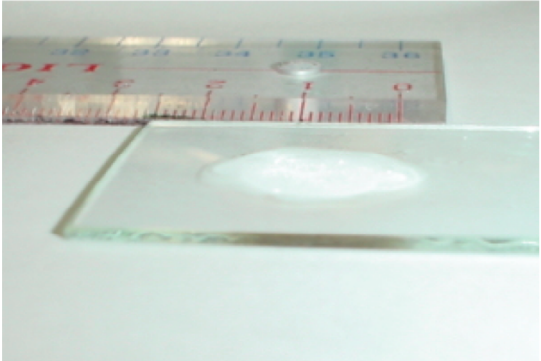
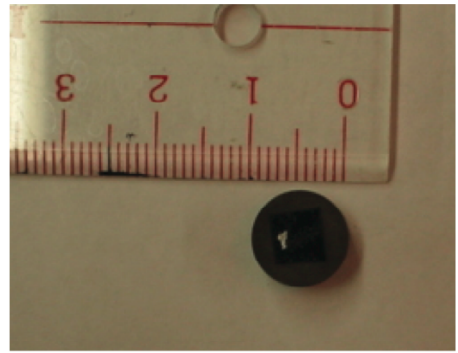

Fig. 6. The setup of the test-explosion experiment done by the JIG. This figure was taken from Ref. [1].

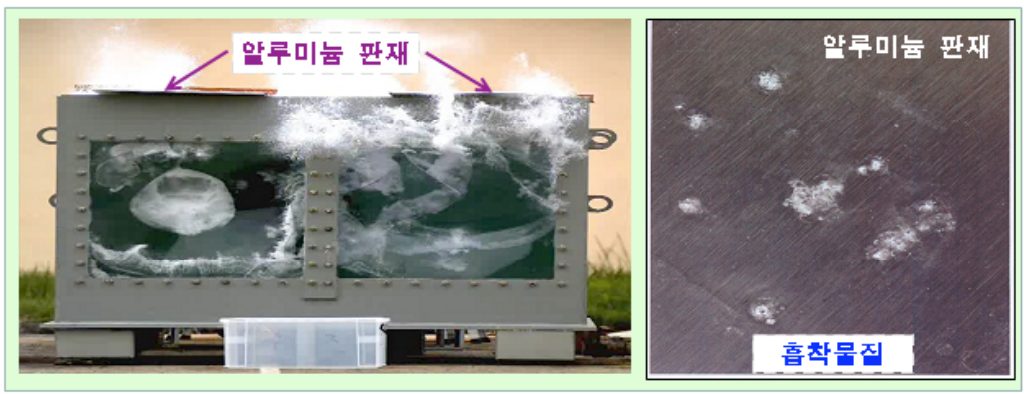